\documentclass[pra,preprint,superscriptaddress,aps,longbibliography]{revtex4-1}
\usepackage{natbib}
\usepackage{graphicx}
\usepackage{enumitem}
\DeclareGraphicsExtensions{.pdf,.png,.jpg}
\usepackage{array}
\usepackage{amsmath}
\usepackage{amsthm}
\usepackage{amssymb}    
\usepackage{latexsym}
\usepackage{amsfonts}
\usepackage{mathrsfs}
\usepackage{color}
\usepackage{bbold}
\usepackage{kotex}
\usepackage{float}
\usepackage{physics}
\newtheorem{theorem}{Theorem}

\usepackage[colorlinks=true, urlcolor=blue,citecolor=blue,linkcolor=blue]{hyperref}
\usepackage{xcolor}
\usepackage[justification=centerlast]{caption} 
\usepackage{subcaption}
\usepackage{tabularx}

\newcommand{\vmu}{\vb*{\mu}}
\newcommand{\vpsi}{\vb*{\psi}}
\newcommand{\veta}{\vb*{\eta}}
\newcommand{\vtheta}{\vb*{\theta}}
\newcommand{\vW}{\vb*{W}}

\begin{document}
\newcommand{\ave}[1]{\left<#1\right>}
\renewcommand{\d}[1]{\ensuremath{\operatorname{d}\!{#1}}}
\renewcommand\qedsymbol{$\blacksquare$}

\title{A variational quantum algorithm by Bayesian Inference with von Mises-Fisher distribution}
\author{Trung Huynh}
\affiliation{Department of Physics, Hanyang University, Seoul, 04763, Republic of Korea}
\author{Gwangil An}
\affiliation{Department of Physics, Hanyang University, Seoul, 04763, Republic of Korea}
\author{Minsu Kim}
\affiliation{Department of Physics, Hanyang University, Seoul, 04763, Republic of Korea}
\author{Yu-Seong Jeon}
\affiliation{Department of Physics, Hanyang University, Seoul, 04763, Republic of Korea}
\author{Jinhyoung Lee}
\email{hyoung@hanyang.ac.kr}
\affiliation{Department of Physics, Hanyang University, Seoul, 04763, Republic of Korea}
\affiliation{Center for Quantum Simulation, Korea Institute of Science and Technology (KIST), Seoul, 02792, Republic of Korea}

\begin{abstract}
	The variational quantum eigensolver algorithm has gained attentions due to its  capability of locating the ground state and ground energy of a Hamiltonian, which is a fundamental task in many physical and chemical problems.  Although it has demonstrated promising results, the use of various types of measurements remains a significant obstacle. Recently, a quantum phase estimation algorithm inspired measurement scheme has been proposed to overcome this issue by introducing an additional ancilla system that is coupled to the primary system. Based on this measurement scheme, we present a novel approach that employs Bayesian inference principles together with von Mises-Fisher distribution and theoretically demonstrates the new algorithm's capability in identifying the ground state with certain for various random Hamiltonian matrices. This also opens a new way for exploring the von Mises-Fisher distribution potential in other quantum information science problems.
\end{abstract}
\maketitle

\section{Introduction} 

Finding the ground state and its energy of a Hamiltonian matrix  is a fundamental task in many physical and chemical processes. Among several methods, the variational quantum eigensolver (VQE) \cite{ch5.Peruzzo.2014} has become prominent in the last decade due to its capability of searching the solution within polynomial time. The main idea of this method is dividing the complex Hamiltonian matrix into smaller parts and using quantum circuit to calculate their expectations, then a classical optimizer is employed to find the new trial probe for the next iteration until the termination conditions are satisfied. One of the key advantages of VQE is its potential for near-term applications, where quantum computers with limited qubit numbers and gate fidelity are available. Several studies have investigated the performance of VQE on different systems and problems  such as small molecules and quantum magnets \cite{ch5.Liu.2019,ch5.Kandala.2017}, computing the electronic transitions \cite{ch5.Parrish.2019}, frustrated quantum systems \cite{ch5. Uvarov.2020}, and dynamic correlation functions \cite{ch5.Chen.2021}. As this method calculates the expectation of the Hamiltonian through direct measurements on the system, one obvious obstacle is that the number of measurement types  needed rapidly increases when the dimension of the system grows up. Though there have been attempts but this issue remains a challenging problem (for an instance, see a review by \cite{Tilly.2022}).

One promising method which can be utilized to find the ground state without dealing with such measurements issue is the quantum phase estimation algorithm (QPEA) \cite{Kitaev.1995, Nielsen.2000}. This method is a key component in many quantum applications such as quantum computing \cite{Shor.1994} and quantum metrology \cite{Smith.2024}. The QPEA also plays a crucial role in finding eigenvalues of Hamiltonians \cite{Zhou.2013}, which is essential for simulating quantum systems and solving problems in quantum chemistry and materials science . One drawback of the original method in use is that it requires many ancilla qubits to obtain a high precision, hence increasing the depth of quantum circuits. To overcome this obstacle, ones can employ the iterative phase estimation algorithm which requires a single control qubit as the ancilla system  but still attain the same degree of accuracy \cite{Dobsicek.2007}. Recently, a simple measurement scheme inspired from this circuit was proposed by Santagati et al. \cite{ch5.Santagati.2018} can approximate the eigenvalues and eigenstates for both ground state and excited states with highly accurate results for several specific complex Hamiltonians. However, they utilized the ansatz whose initial overlap with the true state varies from $0.2$ to $0.8$ in their particular examples, which may not available in practice generally.

In order to explore the efficiency of such measurement scheme in the current problem, we develop a new algorithm which can work without prior information that enables good guesses for the ground state as well as it can be applicable to various Hamiltonian matrices based on the Bayesian inference. Bayesian methods have been proved to be useful (and optimal in many cases) in several quantum information tasks such as quantum tomography \cite{Huszar.2012}, quantum phase estimation \cite{Smith.2024}, quantum circuit learning \cite{Granade.2012}, and others (for examples, see in \cite{Xiao.2021, Garcial.2019, Nolan.2021}). Unlike many other works in which the normal distribution is employed, in this work we introduce the von Mises - Fisher (vMF) distribution to apply in the current task. We aim to explore its potential in quantum information science as this distribution plays a crucial role in directional statistics \cite{ch5.Mardia.1999,Fisher.1993} which shares some similar properties of the quantum state vectors space.

\section{Measurement scheme}

Suppose that an initial known state $|\psi\rangle$ undergoes a unitary evolution $\hat{U} =\exp(-i\hat{H}t)$, where $t$ is the evolution time and the Hamiltonian matrix of interest $\hat{H}$ which can be written in diagonalized form as $\hat{H} = \sum_{n=0}^{d-1}\varepsilon_n |h_n\rangle\langle h_n|$  with a known dimension $d$.  The main task now is how to figure out the ground energy $\varepsilon_0$ and ground state $|h_0\rangle\langle h_0|$ by inferring the data from the measurements on the quantum state. In this work, it is supposed $\varepsilon_nt \in (0,\pi]$ for all $n$. This assumption can be satisfied in general by scaling the original Hamiltonian matrix and choosing the value for the time $t$ given some prior knowledge about energy lower and upper bound \cite{H.scaling}, which may be available in practice. 

\begin{figure}[h!]
	\includegraphics[width=0.6\linewidth]{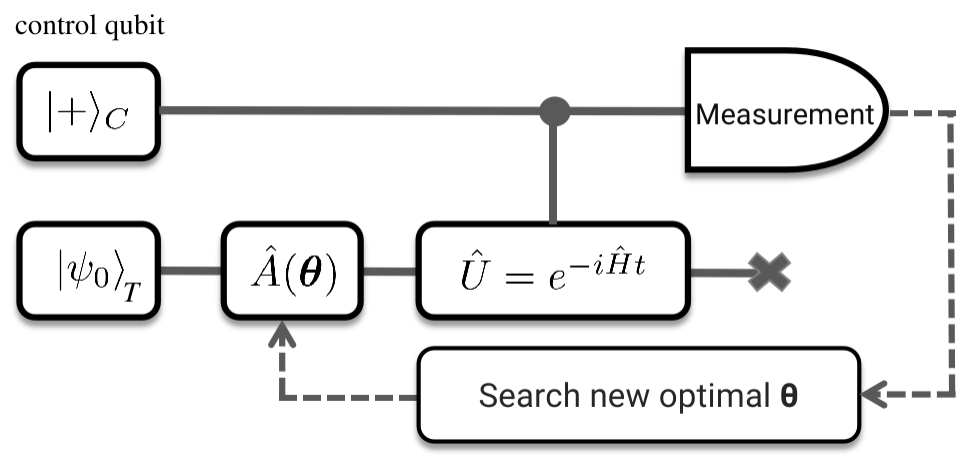}
	\caption[The illustration of the measurement setup.]{The illustration of the measurement setup. We perform the measurements on the control qubit only with the basis $\{|+\rangle, |-\rangle\}$. The outcome $x$ is assigned to "success" ($x=0$) or "fail" $(x=1)$. The operation $\hat{A}(\vb*{\theta})$ acts on a reference state ($|0\rangle$, for example) to generate trial states. }
	\label{fig:vqesetup}
\end{figure}

In the Fig. (\ref{fig:vqesetup}), we outline the specifics of the measurement scheme designed to determine the ground state of the Hamiltonian. In the original VQE measurement approach, measurements are directly conducted on the main system to compute the expectation values $\langle \psi|\hat{H}|\psi\rangle$ \cite{ch5.Peruzzo.2014}. However, this method encounters the required number of measurement types when the system's dimension increases as aforementioned. To avoid this issue, we can borrow the idea from quantum phase estimation algorithm in which making the measurements on the ancilla systems coupled with the main system can extract the information we are desired to know. Instead of employing many control qubits, the ancilla system in our case is a single qubit only initially prepared in the state $|+\rangle$. The interaction between two systems is governed by a control-unitary operation, which is assumed to be perfectly realized in our work. The practical construction of this type of operation can be found in \cite{ch5.Zhou.2011, ch5.Patel.2016}.

Assume that our initial state is $|+\rangle\otimes |\psi\rangle$,  the control-unitary operation encodes the information of the eigenvalues of the Hamiltonian  in the phase of the control qubit:
\begin{equation}
	|\psi_{tot}\rangle=	\hat{U}_{control} (|+\rangle\otimes|\psi\rangle)  = \frac{1}{\sqrt{2}}(|0\rangle\otimes \hat{I}|\psi\rangle+|1\rangle\otimes \hat{U}|\psi\rangle)=\sum_n\frac{c_n}{\sqrt{2}}(|0\rangle +e^{-i\varepsilon_n t}|1\rangle)\otimes|h_n\rangle.
\end{equation}
As we make the measurements on the control qubit only, the reduced state becomes
\begin{align}
	\rho_C =\Tr_T [|\psi_{tot}\rangle\langle \psi_{tot}|]=\frac{1}{2}\left(\mathbb{I}+\hat{\sigma}_x\sum_n\cos\varepsilon_nt |\langle h_n|\psi\rangle|^2-\hat{\sigma}_y\sum_n\sin\varepsilon_nt |\langle h_n|\psi\rangle|^2  \right),
\end{align}
where $\hat{\sigma}_x, \hat{\sigma}_y$ are Pauli matrices. Therefore, we perform a two-outcome measurement on x-y plane and obtain the outcome probability 
\begin{align}
	p(x|\hat{H},|\psi\rangle,\varphi) =\frac{1}{2}\left[ 1+(-1)^x\sum_{n=0}^{d-1}\cos(\varepsilon_nt+\varphi)|\langle h_n|\psi\rangle|^2\right],
\end{align}
where $x=\{0,1\}$ is the measurement outcome, $\varphi$ is the measurement phase. Choosing $\varphi=0$, we derive the simplified outcome probability function
\begin{align}
	p(x|\hat{H},|\psi\rangle) =\frac{1}{2}\left[ 1+(-1)^x\sum_{n=0}^{d-1}\cos\varepsilon_nt|\langle h_n|\psi\rangle|^2\right].\label{outcome_prop}
\end{align}
It is easy to see that if the trial state $|\psi\rangle$ coincides with the ground state $|h_0\rangle$, the probability $p(x=0)$ is maximized.

\section{Bayesian inference}
Bayesian inference is mainly about the process of updating our knowledge of a variable  when we get some new data \cite{Mockus.1989, Garnett.2023}. This is based on the Bayes's theorem for two events $A$ and $B$, which can be found in any statistical textbook:
\begin{align}
	p(A|B) = \frac{P(B|A)P(A)}{P(B)}.
\end{align}
Assume that we want to study a variable $\vb*{\theta}$, which plays an important role in an experiment. For a given $\vtheta$, we can conduct the experiment to observe some data whose relation can be described by a probability distribution $p(D|\vtheta)$, commonly called \textit{likelihood} function. Now let us model our current knowledge about $\vb*{\theta}$ by another distribution, called \textit{prior} distribution, $p_0(\vb*{\theta})$. After receiving the data, our knowledge has been changed and hence the distribution alters to another one, called \textit{posterior} distribution, $p_1(\vb*{\theta}|D)$. The relation between these distributions is
\begin{align}
	p_1(\vb*{\theta}|D)=\frac{1}{z}p(D|\vb*{\theta})p_0(\vb*{\theta}),
\end{align}
where $z = \int d\vb*{\theta}p(D|\vb*{\theta})p_0(\vb*{\theta})$ is the normalization constant which plays a role as $P(D)=P(B)$.  Before doing any new observation, $p_1(\vb*{\theta}|D)$ becomes the new \textit{prior} distribution as it reflects our current knowledge.  Based on this idea, our algorithm will work iteratively by keeping updating the distribution.  

\subsection{State parametrization}
 There are several ways to realize the quantum states in general, depending on the specific situations such as different laboratory devices, nature of the states. Then it may be complicated if we want to apply a specific preparation procedure to another situation where that method is not specially designed for because the mappings between different states parametrization approaches might not exist. This motivates us to seek a way which can be applied to other methods  with more freedom by existing mappings. 
 
The above requirement suggests that a general way to represent the quantum states can be the best candidate for our purpose. To this end, we can express any arbitrary state vector as $|\psi\rangle = \sum_{k=1}^d \psi_k|k\rangle = \sum_{k=1}^d (\psi_k^r+i\psi_k^i)|k\rangle$, where $\psi_k^r, \psi_k^i$ are real numbers for all $k$. First, this way can cover all the arbitrary quantum states as we employ the full Hilbert space, which can be truncated into subspaces if given knowledge about the solution is available. Next, this parametrization method can minimize the number of free parameters. To see this, note that we need $2d = 2 \times 2^{n_q}$ real numbers first. However, we can neglect a parameter to exclude the global phase and another one due to the unity trace. Thus in total we just need $2d - 2 = 2^{n_q+1} - 2$ free parameters, which cannot be reduced further. Finally, there exists protocols \cite{ch5.Mottonon.2004, ch5.Plesch.2011, ch5.Niemann.2016, ch5.XMZhang.2022} to transform this parametrization to other specific parameters sets, especially for generating quantum states on quantum circuits. Employing this approach to prepare quantum states on some quantum simulators such as Qiskit \cite{Qiskit} and PennyLane \cite{PennyLane} is available and easily done via their built-in functions. For this sake of generality, we utilize the real vector $\vpsi = (\psi_1^r,\psi_1^i,\cdots)$ as the quantum state parametrization for the state $|\psi\rangle$ through out this work. 
 
 


The choice of state parametrization above makes a transition from complex vectors $|\psi\rangle$ to real vectors $\vpsi$.  It is thus necessary to find the real representation for quantum expectation in the likelihood function (\ref{outcome_prop}), which can be rewritten (\ref{outcome_prop}) as
\begin{align}
	p(x\big\vert|\psi\rangle) &=\frac{1}{2}\left[1+(-1)^x\sum_n\cos\varepsilon_n|\langle h_n|\psi\rangle|^2\right] \\
	&= \frac{1}{2}[1+(-1)^x\langle \psi|\hat{V}|\psi\rangle ],
\end{align}
where $\hat{V}= \sum_n \cos\varepsilon_nt |h_n\rangle \langle h_n|$. It can be shown that the expectation $\langle \psi|\hat{V}|\psi\rangle$ can be expressed as $\vpsi^T\vW\vpsi$ where $\vW$ is a real $2d \times 2d$ matrix which plays a role the same as $\hat{V}$. Therefore the outcome distribution now becomes
\begin{align}
	p(x|\vpsi) = \frac{1}{2}[1+(-1)^x\vpsi^T\vW\vpsi].
\end{align}
A proof (see appendix \hyperref[appen_A]{A}) shows that all the valid unit vectors satisfy $\vpsi^T\vW\vpsi\leq \cos\varepsilon_0t$, as well as  $||\vW\vpsi||\leq \cos\varepsilon_0t$ for all $\vpsi$. 

\subsection{{Prior distribution}}
 A natural question arises now: which probability distribution should be used for $\vpsi$? Typically, Gaussian distribution is a popular choice for prior distributions, given its widespread applicability in various problems. However, in our case, we are seeking a distribution that is well-suited for unit vectors. Among potential candidates, we have opted for the von Mises-Fisher (vMF) distribution, which is renowned in directional statistics \cite{Fisher.1953, Fisher.1993, ch5.Mardia.1999}.
 
We now briefly introduce some properties of that distribution function. Let vectors $\vb*{\psi}, \vb*{\mu}$ be unit vectors with dimension $p$, the von Mises-Fisher distribution for the general case \cite{ch5.Mardia.1999} is
\begin{align}
	p(\vb*{\psi}| \vb*{\mu}, k) =C_p(k)\exp(k\vmu^T \vb*{\psi}),
\end{align}
where  $k$ is the concentration parameter, $\vmu$ is the mean direction vector, $C_p(k)=\frac{k^{p/2-1}}{(2\pi)^{p/2}I_{p/2-1}(k)}$, $I_p(k)$ is the modified Bessel function of the first kind at order $p$. The concentration parameter $k$ describes how data distribute over the variable space, which can be considered proportional to $1/\sigma$ with standard deviation $\sigma$ in the normal distribution. If the data is uniformly distributed, $k$ is zero  whereas it is much larger than 1 in the case of localized data.  The expectation of $\vb*{\psi}$ is $\mathbb{E}[\vb*{\psi}|p(\vb*{\psi}|\vb*{\mu}, k)] = A_p(k)\vb*{\mu} = \frac{I_{p/2}(k)}{I_{p/2-1}(k)}\vb*{\mu}$. A more detail on how to find the moment of all orders can be found in appendices \hyperref[first_moment_vMF]{B} and \hyperref[higher_moments_vMF]{C}. 

There are three advantages of the vMF distribution that is useful for our work. First, its form is simple. In any dimension, the function is fully described by a mean vector and scalar concentration parameter while the normal distribution requires a vector and a matrix. The simplicity may reduce the amount of computation and calculation complexity when the dimension grows up. Second, as this function deals with unit vectors only, we do not need any normalization step as if we employ other distribution whose vector's length is arbitrary and hence we can avoid issues such as two different vectors represent the same quantum state, or null vector. Other advantage of using the vMF distribution in practice is that the sampling  can be done by using the inverse method without sample rejection \cite{ch5.Kurz.2015}, preventing us from throwing unwanted samples away. 
\begin{figure}[H]
	\centering
	\includegraphics[width=0.5\linewidth]{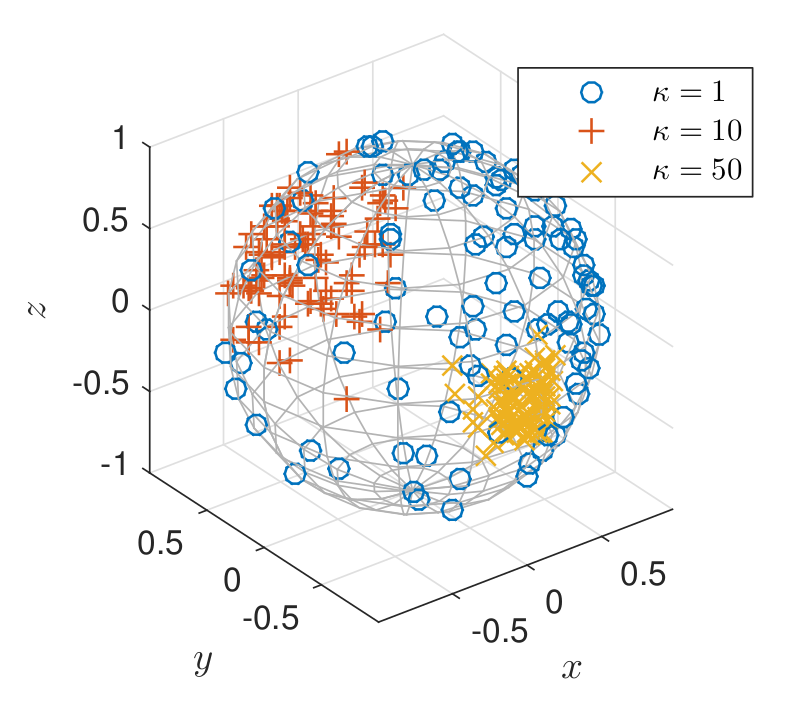}
	\caption[Samples drawn from a vMF distribution with few concentration parameter values ]{Samples drawn from a vMF distribution with few concentration parameter values on the sphere ($p=3$). The figure is taken from \cite{ch5.Kurz.2015}.}
	\label{fig:ch5.vMF}
\end{figure}
Given $N$ independent identical distributed unit vectors $\vpsi_i$ drawn from a vMF distribution, the maximum likelihood estimate of the mean direction  is $\hat{\vmu} = {\bar{\vpsi}}/{R},$ where $\bar{\vpsi}=\frac{1}{N}\sum_i\vpsi_i$ and $R=||\bar{\vpsi}||$ is the resultant length \cite{ch5.Mardia.1999}. The concentration parameter $k$ can be approximated \cite{ch5.Sra.2011} by  $\hat{k}= {R(p-R^2)}/{(1-R^2)}.$ 

\subsection{{Bayesian updating}}

Equipped with all the tools above, now we discuss how the algorithm works. Suppose we are at \textit{n}th iteration, our knowledge is wrapped in a prior distribution $p_n(\vpsi)$. We then generate some trial states from $p_n(\vpsi)$ and do measurements on them to get some data. Based on the new data, we can find the new posterior $p_{n+1}(\vpsi)$. As this new distribution is supposed to be the prior distribution in the next iteration, we need to find its mean vector and concentration parameter. If we consider a success outcome only, the posterior distribution $p_{n+1}(\vpsi|x = 0)$  now is
\begin{align}
	p_{n+1}(\vpsi|x=0) = \frac{1}{2z_n}(1+ \vpsi^T\vW\vpsi)p_n(\vpsi),
\end{align}
where $z_n=\int d\vpsi p(x=0|\vpsi)p_n(\vpsi)$. The reason why we choose $x=0$ only is that it is a sign of finding the ground state. For a given state, as we said, the success probability $p(x=0|\vpsi)$ gets maximal when $\vpsi $ coincides with the ground state. This function adjusts the value of $p_{n+1}(\vpsi)$ compared with $p_{n}(\vpsi)$. If the value of the likelihood function is large, it is likely the posterior value may be larger than of the prior and vice versa. After many iterations, we expect that the posterior distribution will be localized to a small region where the ground state stays in while its values at other regions almost go to zero. Therefore in each iteration, finding the new mean vector and concentration parameter is the central work. 

Denote $\mathbb{E}_{n}[f(\vpsi)]=\int d\vpsi f(\vpsi) p_n(\vpsi)$, we can prove that (see appendix \hyperref[higher_moments_vMF]{C})
\begin{align}
	\mathbb{E}_{n+1}[\vpsi] =\int d\vpsi \vpsi p_{n+1}(\vpsi)&= \frac{1}{2z_n}\int (1+\vpsi^T\vW\vpsi)\vpsi p_n(\vpsi)\\
	&=\frac{\alpha_n}{2z_n}\mathbb{E}_n[\vpsi]+\frac{\beta_n}{2z_n}\vW\mathbb{E}_n[\vpsi],\label{Epsi1Epsi2}
\end{align}
where the coefficients $\alpha_n = 1+\frac{1}{R_n}\left(\frac{B_n}{k_n}\Tr\vW + D_n\vmu_n^T\vW\vmu_n\right),	\beta_n= \frac{2B_n}{R_nk_n}, R_n = ||\mathbb{E}_{n}[\vpsi]||,	B_n =\frac{I_{p/2+1}(k_n)}{I_{p/2-1}(k_n)}, 	D_n =\frac{I_{p/2+2}(k_n)}{I_{p/2-1}(k_n)}$ are non-negative numbers. The new mean vector is thus given by
\begin{align}
	\vmu_{n+1} = \frac{\mathbb{E}_{n+1}[\vpsi]}{||\mathbb{E}_{n+1}[\vpsi]||} = \frac{\alpha_n \vmu_n +\beta_n \vW\vmu_n}{||\alpha_n \vmu_n +\beta_n \vW\vmu_n||}.\label{mu1mu0}
\end{align}
The new concentration parameter $k_{n+1}$ for the next iteration can be found by the estimator defined above with $R_{n+1} = ||\mathbb{E}_{n+1}[\vpsi]||$.

\section{Convergence of the solution}

The most crucial question for this sort of algorithm  is that whether it can find the ground state (supposed to be $\vmu_t$) or not. Our algorithm's solution converges if the new mean vector is closer to the true solution than that of previous mean vectors.

\begin{theorem}
At any \textit{n}th iteration, the overlap between the new mean vector $\vmu_{n+1}$ to the true vector $\vmu_t$ is always equal to or larger than that of the current mean vector $\vmu_n$.
\end{theorem}
In order to prove the theorem, the following inequality must be hold in any case:
\begin{align}
	R_c = \frac{\vmu_t^T\vmu_{n+1}}{\vmu_t^T\vmu_n} \geq 1.\label{ch5.converge.ratio}
\end{align}
We employ the relation (\ref{mu1mu0}) and $R_c \geq 1$ if
\begin{align}
	{\alpha_n+\beta_n\frac{\vmu_t^T\vW\vmu_n}{\vmu_t^T\vmu_n}}\geq{ || \alpha_n \vmu_n +\beta_n \vW\vmu_n   ||  } .
\end{align}
It can be shown that $\vmu_t^T\vW = \cos\varepsilon_0t\vmu_t^T$ (see the appendix \hyperref[appen_A]{A}), and the RHS is always upper bounded by $\alpha_n + \beta_n ||\vW\vmu_n||$, therefore the condition now becomes $||\vW\vmu_n|| \leq \cos\varepsilon_0t,$ which is always hold for any $\vmu_n$ as we said before. It is worthy noticed that the equality just happens when the current mean vector is the ground state only, hence the ratio is strictly greater than $1$ in any case when $\vmu_n \neq \vmu_t$. In other words, the algorithm's solution always converges.

In practice, it's common to search for solutions within a subspace of parameters rather than the entire parameter space. This limitation arises due to experimental constraints, such as a limited number of measurements or prior knowledge about the region where the true solution is likely to be found. If the true solution $\vmu_t$ resides within this subspace, the convergence ratio (\ref{ch5.converge.ratio}) still remains valid. Conversely, if the true solution lies outside the subspace, the solution will converge to a state $\tilde{\vmu_t}$ where $\tilde{\vmu_t}$ represents the vector within the subspace that best aligns with $\vmu_t$.  Therefore, searching in the subspace also assures the convergence of the solution.

\begin{theorem}
When the number of iteration increases, not only the mean vector approaches to the true solution, but  its  resultant length $||\mathbb{E}[\vpsi]||$ also converges. 
\end{theorem}

To see that, let us recall $\mathbb{E}_{n+1}[\vpsi] = \frac{||\mathbb{E}_n[\vpsi]||}{2z_n}(\alpha_n+\beta_n\vW)\vmu_n$. Because the concentration parameter $k_{n}$ is monotonic with the mean vector length \cite{ch5.Mardia.1999, ch5.Sra.2011}, $k_{n+1}$ gets larger than or equal to $k_n$ if $||\mathbb{E}_{n+1}[\vpsi]|| \geq ||\mathbb{E}_n[\vpsi]||$, or $||(\alpha_n+\beta_n\vW)\vmu_n|| \geq 2z_n$. 

First, we note that 
\begin{align}
	|| (\alpha+\beta_n\vW)\vmu_n||^2=\vmu_n^T(\alpha_n+\beta_n\vW)^2\vmu_n&\geq [\vmu_n^T(\alpha_n+\beta_n\vW)\vmu_n]^2=(\alpha_n+\beta_n\vmu_n^T\vW\vmu_n)^2.
\end{align}
Thus we need $\alpha_n+\beta_n\vmu_n^T\vW\vmu_n \geq 2z_n$. Let us consider the coefficients in  more details:
\begin{gather}
	\alpha_n = 1 +\frac{B_n}{A_nk_n}\Tr\vW+\frac{D_n}{A_n}\vmu_n^T\vW\vmu_n,\quad\quad	\beta_n\vmu_n^T\vW\vmu_n = \frac{2B_n}{A_n k_n}\vmu_n^T\vW\vmu_n, \\
	2z_n = 1+\frac{A_n}{k_n}\Tr\vW + B_n\vmu_n^T\vW\vmu_n,
\end{gather}
hence we require
\begin{align}
	\left(\frac{B_n}{A_n}-A_n\right)\frac{1}{k_n}\Tr\vW &+\left[ \frac{1}{A_n}\left(D_n+\frac{2B_n}{k_n}\right)-B_n\right]\vmu_n^T\vW\vmu_n \geq 0\\
	\Leftrightarrow (A_n^2-B_n)\frac{1}{k_n}\Tr\vW & \leq\left(D_n+\frac{2B_n}{k_n}-A_nB_n\right)\vmu_n^T\vW\vmu_n.
\end{align}
Now remind that $A_n = \frac{I_v(k_n)}{I_{v-1}(k_n)}, B_n = \frac{I_{v+1}(k_n)}{I_{v-1}(k_n)}, D_n= \frac{I_{v+2}(k_n)}{I_{v-1}(k_n)}$ with $v=p/2$. After some calculations, we can obtain $A_n^2\geq B_n$ and $\frac{k_n(D_n+2B_n/k_n-A_nB_n)}{A_n^2-B_n} = 2v =p$. This implies that the concentration parameter converges if
\begin{align}
	\vmu_n^T\vW\vmu_n \geq \frac{1}{p}\Tr\vW. \label{ch5.kappa.condition1}
\end{align}
The left hand side, by previous proof, will converge to the value of $\cos\varepsilon_0 t$ while the right hand side is the average $\frac{1}{p}\sum_{n=0}^{d-1}\cos\varepsilon_nt$, which is smaller than the left hand side. Consequently, in the long run, the concentration increases or the resultant length $R_n \rightarrow 1$. 


In the following figures, we plot some examples with theoretical calculations based on the theory. We examine two prototypes such as Helium Hydride ion \textbf{He-H}$^+$ and Hydrogen molecule \textbf{H}$_2$ and compute the fidelity as well as the resultant length $R_s$ over the iterations. The initial guesses are chosen randomly and the algorithm runs with the exact value of the matrix $\vW$ within fixed number of iterations. For each prototype, we repeat the algorithm with $100$ different initial guesses. Therefore, in each iteration, all the quantities are averaged from $100$ runs. The initial value for the concentration parameter $k=0.001$. The algorithm stops when either the number of iteration  or $k$ attains a threshold value. In particular, we choose $k_{max} = 700$ and $n^{iteration}_{max} = 1000$.

\begin{figure}[H]
	\centering
	\includegraphics[width=0.8\linewidth]{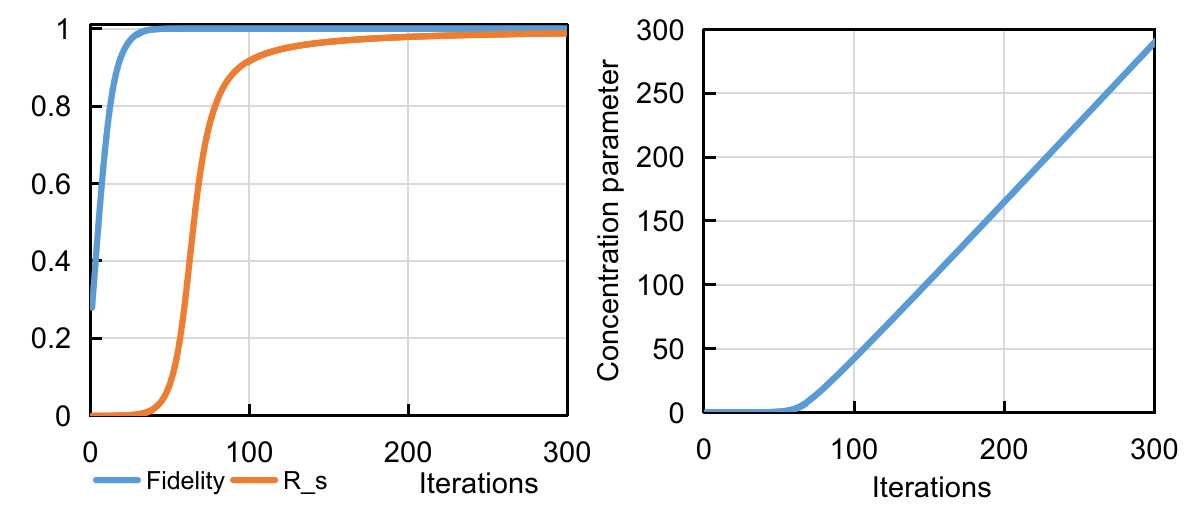}
	\caption[Theoretical predictions for fidelity, mean vector length (a) and concentration parameter (b) for 1 qubit.]{Theoretical predictions for fidelity, mean vector  resultant length (left) and concentration parameter (right) for \textbf{He-H}$^+$, which is a prototype examined in original variational quantum eigensolver paper \cite{ch5.Peruzzo.2014}.}
	\label{fig:theory1qubit}
\end{figure}

\begin{figure}[H]
	\centering
	\includegraphics[width=0.8\linewidth]{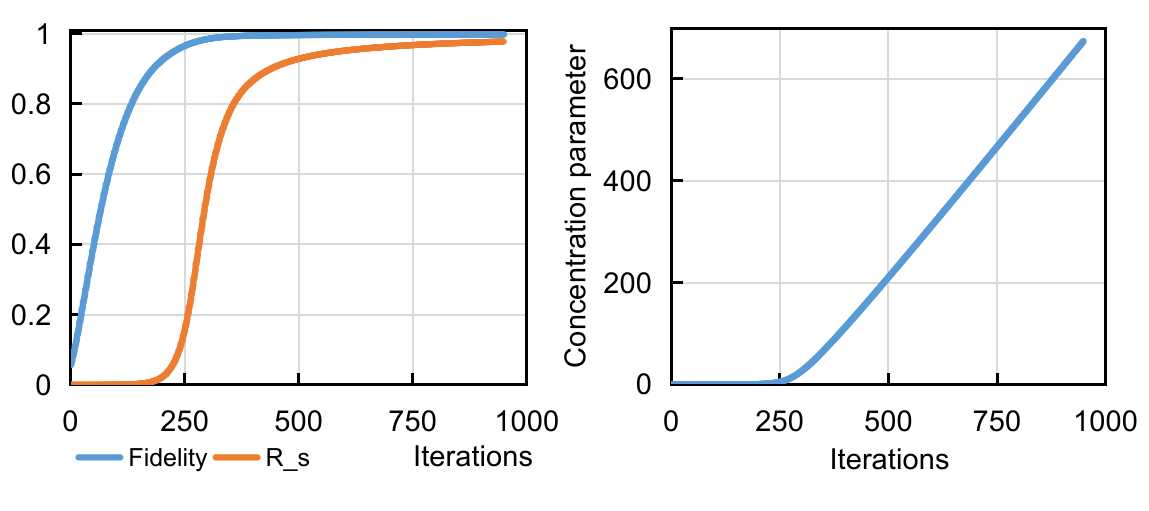}
	\caption[]{Results for the Hamiltonian of the Hydrogen Molecule (6-31G basis) set at an interatomic distance of 0.745 $\mathring{A}$, see \cite{Shee.2022} for the explicit form. It is clear that we need more iterations to obtain the convergence when the dimension of the system is larger.} 
	\label{fig:theory23qubit}
\end{figure}

It can be seen that results confirm all the predictions of the theory we propose above. This stimulates us to design a corresponding practical algorithm to apply to the real situations when the information of the Hamiltonian matrix $\hat{H}$ (or $\vW$) is incomplete, which will be shown in another paper in which we discuss in details many aspects of realizing that practical algorithm. 

\section{Remarks}

The variational quantum eigensolver has emerged as a promising approach for determining the ground state and ground energy of Hamiltonian matrices, critical for addressing various physics and chemistry challenges. However, it faces several challenges, particularly in terms of the numerous types of measurements required. Addressing this, a quantum phase estimation algorithm variant scheme \cite{ch5.Santagati.2018} has shown promising results for some specific physical systems. The main idea is that the information of the Hamiltonian can be transferred to an ancilla single qubit coupled with the main system. Though the dimension of the system increases, just a two-outcome measurement setting only is needed to infer the encoded information, which is key advantage of this scheme.
In our study, we employ this scheme and introduce a Bayesian inference theory utilizing the von Mises-Fisher distribution to achieve our objectives. Our theoretical analysis has shown the capability of our algorithm in finding solutions for any Hamiltonian matrix with eigenvalues assumed within the range $(0, \pi]$ in the long term without requiring any prior knowledge about the solution. The convergence of the solution is not only guaranteed when searching the whole Hilbert space but also applied for the subspaces, which is promising both theoretically and practically. This lays the groundwork for the development of a practical algorithm which will be presented in a forthcoming paper. Finally, the combination between Bayesian inference and vMF distribution has been illustrated a good optimizer for our work, which may be extended to other circumstances in which unit vectors are involved. Our exploration of the von Mises-Fisher distribution's application also opens up a new road for its study in other quantum information science problems.

\begin{acknowledgments}
JL was supported by the National Research Foundation of Korea (NRF) grant (No. 2022M3E4A1077369 and NRF-2023M3K5A1092818) funded by the Korea government (MSIT).
\end{acknowledgments}

\appendix

\section*{Appendices}

\subsection*{A. Real representation of quantum expectation}\label{appen_A}
Let us express $|\psi\rangle=\sum_{k}(\psi^r_k+i\psi^i_k)|k\rangle$, and $|h\rangle = \sum_{k}(h^r_k+ih^i_k)|k\rangle$, we can compute the inner product 
\begin{align}
	\langle h|\psi\rangle &= \sum_{j,k}(h^r_j-ih^i_j)(\psi^r_k+i\psi^i_k)\langle j|k\rangle =\vb*{h}^T \vpsi+i(\vb*{h}^s)^T \vpsi,
\end{align}
where $\vb*{h}=[h_1^r,h_1^i,h_2^r,h_2^i,\cdots]^T$, $\vb*{h}^s=[-h_1^i,h_1^r,-h_2^i,h_2^r,\cdots]^T$ and note that $\vb*{h}^T \vb*{h}^s=0$. Thus we have
\begin{align}
	\langle\psi|\hat{V}|\psi\rangle &= \sum_nc_n\langle \psi|h_n\rangle \langle h_n|\psi\rangle=\sum_nc_n(\vpsi^T \vb*{h}_n-i\vpsi^T \vb*{h}_n^s)(\vb*{h}_n^T \vpsi+i(\vb*{h}_n^s)^T \vpsi)\\
	&=\sum_nc_n\vpsi^T [\vb*{h}_n\vb*{h}_n^T+\vb*{h}_n^s(\vb*{h}_n^s)^T] \vpsi=\vpsi^T \vW \vpsi,
\end{align}
where $\vW= \sum_nc_n[\vb*{h}_n\vb*{h}_n^T+\vb*{h}_n^s(\vb*{h}_n^s)^T]$. For each quantum state $|h_k\rangle$, there are two corresponding real vectors $\vb{h}_k$ and $\vb{h}_k^s$. The orthogonality now requires not only $\vb*{h}_j^T \vb*{h}_k =\delta_{j,k}$ but also $(\vb*{h}_j^s)^T \vb*{h}_k=0,$ for all $j,k$.

Another notice is that $\vW^2=\sum_n c_n^2[\vb*{h}_n\vb*{h}_n^T+\vb*{h}_n^s(\vb*{h}_n^s)^T]$, then we have
\begin{align}
	\vpsi^T\vW^2\vpsi &= \sum_n c_n^2\vpsi^T[\vb*{h}_n\vb*{h}_n^T+\vb*{h}_n^s(\vb*{h}_n^s)^T]\vpsi=\sum_n c_n^2 \langle \psi |h_n\rangle \langle h_n|\psi\rangle\\
	&\leq c_0^2\sum_n \langle \psi |h_n\rangle \langle h_n|\psi\rangle=c_0^2\langle \psi | \left(\sum_n |h_n\rangle \langle h_n|\right) |\psi\rangle = c_0^2\\
	\Rightarrow |\vW\vpsi| &\leq c_0=\cos\varepsilon_0t.
\end{align}
This is the property we used to prove the convergence of the solution.

\subsection*{B. Calculation of the first moment of vMF distribution }\label{first_moment_vMF}

First, we express the unit vector $\vb*{\mu} $ in terms of a non-unit vector $\vb*{\eta}$ and the summation to unity reads:
\begin{gather}
	C_p(k)\int d\vb*{\psi} \exp\left(k\frac{\veta^T}{||\veta||} \vpsi\right)=1\label{eq_sum_1}\\
	\Rightarrow \frac{\partial }{\partial \vb*{\eta}}  C_p(k)\int d\vb*{\psi}\exp\left(k\frac{\veta^T}{||\veta||} \vpsi\right) =\vb*{0}\\
	\Leftrightarrow C_p(k)\int d\vb*{\psi} \left(\frac{k\vb*{\psi}}{||\vb*{\eta}||}-\frac{k\veta^T \vpsi}{||\vb*{\eta}||^2}\frac{\partial\ ||\vb*{\eta}||}{\partial\vb*{\eta}}\right)\exp\left(k\frac{\veta^T}{||\veta||} \vpsi\right)=\vb*{0}\\
	\Leftrightarrow \frac{k}{||\vb*{\eta}||}\mathbb{E}[\vb*{\psi}] = \frac{k}{||\vb*{\eta}||^2}\frac{\partial\ ||\vb*{\eta}||}{\partial\vb*{\eta}}\vb*{\eta}^T \mathbb{E}[\vb*{\psi}]\\
	\Leftrightarrow \mathbb{E}[\vpsi] = \frac{\partial ||\veta||}{\partial \veta} \frac{\veta^T}{||\veta||} \mathbb{E}[\vpsi].\label{expect_1}
\end{gather}
We need to find $\partial ||\vb*{\eta}||/\partial\vb*{\eta}$ and $\frac{\veta^T}{||\veta||} \mathbb{E}[\vpsi]$.  The first term is simply $\partial ||\vb*{\eta}||/\partial\vb*{\eta}= \vb*{\eta}/||\vb*{\eta}||=\vmu$. To find the latter one, we take the derivative with respect to $k$ on the both sides of (\ref{eq_sum_1}):
\begin{gather}
	\frac{\partial }{\partial k} \left[ C_p(k)\int d\vb*{\psi} \exp\left(k\frac{\veta^T}{||\veta||} \vpsi\right)\right]=0\\
	\Rightarrow \frac{1}{C_p(k)}\frac{\partial C_p(k)}{\partial k} C_p(k)\int d\vb*{\psi}\exp\left(k\frac{\veta^T}{||\veta||} \vpsi\right)+ C_p(k)\int d\vb*{\psi} \frac{\veta^T}{||\veta||} \vpsi \exp\left(k\frac{\veta^T}{||\veta||} \vpsi\right)=0\\
	\Leftrightarrow \frac{\veta^T}{||\veta||} \mathbb{E}[\vb*{\psi}] = - \frac{\partial \ln C_p(k)}{\partial k}.
\end{gather}
Therefore, we obtain from (\ref{expect_1}):
\begin{align}
	\mathbb{E}[\vb*{\psi}] = \left(\frac{\veta^T}{||\veta||} \mathbb{E}[\vpsi]\right)  \vb*{\mu} = - \frac{\partial \ln C_p(k)}{\partial k}\vb*{\mu}.
\end{align}
Using $C_p(k)=\frac{k^{p/2-1}}{(2\pi)^{p/2}I_{p/2-1}(k)}$, we have
\begin{align}
	\frac{\partial \ln C_p(k)}{\partial k} &= \frac{\partial }{\partial k}\left[ \left(\frac{p}{2}-1\right)\ln k - \frac{p}{2}\ln 2\pi - \ln I_{p/2-1}(k) \right]\\
	&= \left(\frac{p}{2}-1\right)\frac{1}{k} - \frac{1}{I_{p/2-1}(k)}\frac{\partial I_{p/2-1}(k)}{\partial k}\\
	&= - \frac{I_{p/2}(k)}{I_{p/2-1}(k)},
\end{align}
where we used $\frac{\partial I_v(z)}{\partial z} = \frac{v}{z}I_v(z)+I_{v+1}(z)$, hence
\begin{align}
	\mathbb{E}[\vb*{\psi}] = \frac{I_{p/2}(k)}{I_{p/2-1}(k)}\vb*{\mu}= A_p(k)\vb*{\mu}. 
\end{align}

\subsection*{C. Calculation of the higher moments of vMF distribution }\label{higher_moments_vMF}

In this section, we aim to calculate $\mathbb{E}[\vpsi^{\otimes n}]$, where $\vpsi^{\otimes n}=\vpsi\vpsi\cdots \vpsi$ is the tensor product of $n$ vectors $\vpsi$. Let us start with
\begin{align}
	\mathbb{E}[\vpsi^{\otimes (n-1)}]&=C_p(k)\int d\vpsi \exp(k\vmu^T \vpsi)\vpsi^{\otimes(n-1)}\\
	&=C_p(k)\int d\vpsi \exp(k\frac{\veta^T}{||\veta||} \vpsi)\vpsi^{\otimes(n-1)}\label{psi_n_expectation}\\
	\Rightarrow \frac{\partial}{\partial{\veta}}\mathbb{E}[\vpsi^{\otimes (n-1)}]&=C_p(k)\int d\vpsi \left(\frac{k\vpsi}{||\veta||}-\frac{k\veta^T \vpsi}{||\veta||^2}\frac{\veta}{||\veta||}\right)\exp(k\vpsi \vmu)\vpsi^{\otimes(n-1)}\\
	&=\frac{k}{||\veta||}\mathbb{E}[\vpsi^{\otimes n}]-\frac{k\veta}{||\veta||^2}\frac{\veta^T}{||\veta||} \mathbb{E}[\vpsi^{\otimes n}]\\
	\Rightarrow\mathbb{E}[\vpsi^{\otimes n}]&=\frac{||\veta||}{k}\frac{\partial}{\partial{\veta}}\mathbb{E}[\vpsi^{\otimes (n-1)}]+\frac{\veta}{||\veta||}\frac{\veta^T}{||\veta||} \mathbb{E}[\vpsi^{\otimes n}].
\end{align}
To find the second term in the right hand side, we take the derivative with respect to $k$ on both sides of equation (\ref{psi_n_expectation}):
\begin{align}
	\frac{\partial}{\partial {k}}\mathbb{E}[\vpsi^{\otimes(n-1)}]&=\frac{\partial\ln C_p(k)}{\partial {k}}\mathbb{E}[\vpsi^{\otimes(n-1)}]+\frac{\veta^T}{||\veta||} \mathbb{E}[\vpsi^{\otimes n}]\\
	\Rightarrow \frac{\veta^T}{||\veta||} \mathbb{E}[\vpsi^{\otimes n}]&= \frac{\partial}{\partial {k}}\mathbb{E}[\vpsi^{\otimes(n-1)}] - \frac{\partial\ln C_p(k)}{\partial {k}}\mathbb{E}[\vpsi^{\otimes(n-1)}]\\
	&=\left[\frac{\partial}{\partial{k}}+A_p(k)\right]\mathbb{E}[\vpsi^{\otimes (n-1)}],
\end{align}
where $A_p(k)=-\partial_{k} \ln C_p(k)$. Finally, we get
\begin{align}
	\mathbb{E}[\vpsi^{\otimes n}] = \left\{\frac{||\veta||}{k}\frac{\partial}{\partial{\veta}} +\frac{\veta}{||\veta||}\left[\frac{\partial}{\partial{k}}+A_p(k)\right] \right\}\mathbb{E}[\vpsi^{\otimes (n-1)}].
\end{align}

Now, we want to find $\mathbb{E}[\vpsi\vpsi\vpsi]$ given $\mathbb{E}[\vpsi]=A_p(k)\vmu$. First, we find $\mathbb{E}[\vpsi\vpsi]$ by considering the element $\mathbb{E}[\psi_{\alpha}\psi_\beta]$:
\begin{align}
	\mathbb{E}[\psi_{\alpha}\psi_\beta]&=\left\{\frac{||\veta||}{k}\frac{\partial}{\partial{\eta}_\alpha} +\frac{\eta_\alpha}{||\veta||}\left[\frac{\partial}{\partial{k}}+A_p(k)\right] \right\} A_p(k)\frac{\eta_\beta}{||\veta||}\\
	&=\frac{A_p(k)}{k}\delta_{\alpha,\beta}+\left[A^2_p(k)+\partial_{k}A_p(k)-\frac{A_p(k)}{k}\right]\frac{\eta_\alpha\eta_\beta}{||\veta||^2}\\
	&=\frac{A}{k}\delta_{\alpha,\beta}+B\frac{\eta_\alpha\eta_\beta}{||\veta||^2},
\end{align}
where $A$ is $A_p(k)$, and $B= A^2_p(k)+\partial_{k}A_p(k)-\frac{A_p(k)}{k} =I_{p/2+1}(k)/I_{p/2-1}(k)$. Definitely, $A,B\geq0$. 

Now we can find $\mathbb{E}[\psi_\alpha\psi_\beta\psi_\gamma]$:
\begin{align}
	\mathbb{E}[\psi_\alpha\psi_\beta\psi_\gamma]&=\left[\frac{||\veta||}{k}\frac{\partial}{\partial{\eta}_\alpha} +\frac{\eta_\alpha}{||\veta||}\left(\frac{\partial}{\partial{k}}+A\right) \right] \mathbb{E}[\psi_\beta\psi_\gamma]\\
	&=\left[\frac{||\veta||}{k}\frac{\partial}{\partial{\eta}_\alpha} +\frac{\eta_\alpha}{||\veta||}\left(\frac{\partial}{\partial{k}}+A\right) \right]\left( \frac{A}{k}\delta_{\beta,\gamma}+B\frac{\eta_\beta\eta_\gamma}{||\veta||^2}   \right)\\
	&=\frac{B}{k}\left(\frac{\eta_\alpha}{||\veta||}\delta_{\beta,\gamma}+\frac{\eta_\beta}{||\veta||}\delta_{\alpha,\gamma}+\frac{\eta_\gamma}{||\veta||}\delta_{\alpha,\beta}\right)+\left(AB+\partial_{k}B-\frac{2B}{k}\right)\frac{\eta_\alpha\eta_\beta\eta_\gamma}{||\veta||^3}\\
	&=\frac{B}{k}\left(\frac{\eta_\alpha}{||\veta||}\delta_{\beta,\gamma}+\frac{\eta_\beta}{||\veta||}\delta_{\alpha,\gamma}+\frac{\eta_\gamma}{||\veta||}\delta_{\alpha,\beta}\right)+D\frac{\eta_\alpha\eta_\beta\eta_\gamma}{||\veta||^3},
\end{align}
where $D=AB+\partial_{k}B-\frac{2B}{k}=I_{p/2+2}(k)/I_{p/2-1}(k)$, which is non-negative also.

Next, we can find the term $\mathbb{E}[(\vpsi^T\vW\vpsi)\psi_\gamma]=\sum_{\alpha,\beta}\mathbb{E}[\psi_\alpha W_{\alpha,\beta}\psi_\beta\psi_\gamma]$:
\begin{align}
	\sum_{\alpha,\beta}\mathbb{E}[\psi_\alpha W_{\alpha,\beta}\psi_\beta\psi_\gamma]&=\frac{B}{k}\sum_{\alpha,\beta}W_{\alpha,\beta}\left(\frac{\eta_\alpha}{||\veta||}\delta_{\beta,\gamma}+\frac{\eta_\beta}{||\veta||}\delta_{\alpha,\gamma}+\frac{\eta_\gamma}{||\veta||}\delta_{\alpha,\beta}\right)+D\sum_{\alpha,\beta}\frac{\eta_\alpha W_{\alpha\beta}\eta_\beta\eta_\gamma}{||\veta||^3}
\end{align}
hence,
\begin{align}
	\mathbb{E}_n[(\vpsi^T\vW\vpsi)\vpsi]  &=\frac{2B_n}{k_n}\vW\vmu_n+\left(\frac{B_n}{k_n}Tr\vW+D_n\vmu_n^T\vW\vmu_n\right)\vmu_n\\
	\Rightarrow \mathbb{E}_{n+1}[\vpsi]&=\frac{1}{2z_n}\{\mathbb{E}_n[\vpsi]+\mathbb{E}_n[(\vpsi^T\vW\vpsi)\vpsi]\}\\
	&=\frac{1}{2z_n}\left[ R_n\vmu_n+\frac{2B_n}{k}\vW\vmu_n+\left(\frac{B_n}{k}Tr\vW+D_n\vmu_n^T\vW\vmu_n\right)\vmu_n     \right]\\
	&=\frac{R_n}{2z_n}\left\{ \left[1+\frac{1}{R_n} \left(\frac{B_n}{k}Tr\vW+D_n\vmu_n^T\vW\vmu_n\right)  \right] \vmu_n +\frac{2B_n}{R_nk_n}\vW\vmu_n \right\}\\
	&=\frac{R_n}{2z_n}(\alpha \vmu_n +\beta\vW\vmu_n ),
\end{align}
where $R_n=||\mathbb{E}_n[\vpsi]||$. Finally, we arrive at
\begin{align}
	\vmu_{n+1}=\frac{\mathbb{E}_{n+1}[\vpsi]}{||\mathbb{E}_{n+1}[\vpsi]||} &= \frac{    \alpha_n \vmu_n +\beta_n\vW\vmu_n  }{||\alpha \vmu_n +\beta\vW\vmu_n ||},
\end{align}
where $\alpha_n = 1+\frac{1}{R_n} \left(\frac{B_n}{k_n}Tr\vW+D_n\vmu_n^T\vW\vmu_n\right)$ and $\beta_n=\frac{2B_n}{R_nk_n}$. Obviously, $\alpha_n$ and $\beta_n$ are non-negative numbers.


\begin{thebibliography}{9}
\bibitem{H.scaling} Suppose $\varepsilon_{low}$ and $\varepsilon_{up}$ are the lower and upper bounds of the energies. In order to scale all the eigenvalues of the Hamiltonian $\hat{H}$ into $(0,\pi]$, we can choose the time $t=\pi/\varepsilon_{up}$ if all the bounds are positive. If the lower bound is negative, we can shift the Hamiltonian by $\lambda \hat{I}$ with $\lambda \geq |\varepsilon_{low}|$ and identity matrix $\hat{I}$, then we choose $t = \pi/(\lambda +\varepsilon_{up})$. At the end, we can do some inversion to return the estimate values in the original range. A method for finding the energy bounds can be found in H. Wolkowicz and G. P. H. Styan, Bounds for Eigenvalues Using Traces, Linear algebra and its applications 29:471-506, (1980).
\end{thebibliography}
\end{document}